\documentclass[useAMS,usenatbib]{mn2e}

\usepackage{graphicx}
\usepackage{amssymb}
\usepackage{natbib}
\usepackage{array}
\usepackage{amsmath}
\usepackage{subfigure}
\usepackage[usenames]{color}
\bibpunct{(}{)}{;}{a}{}{,}

\begin{document}

\title[WASP-25b: a 0.6 $M_J$ planet in the Southern hemisphere.]{WASP-25b: a 0.6 $M_J$ planet in the Southern hemisphere.}

\author[B.Enoch et al]
{
B.Enoch$^{1}$\thanks{E-mail:becky.enoch@st-andrews.ac.uk}, 
A.Collier Cameron$^{1}$, 
D.R.Anderson$^{2}$, 
T.A.Lister$^{3}$, 
C.Hellier$^{2}$,
P.F.L.Maxted$^{2}$, 
\newauthor
D.Queloz$^{4}$, 
B.Smalley$^{2}$, 
A.H.M.J.Triaud$^{4}$, 
R.G.West$^{5}$, 
D.J.A.Brown$^{1}$, 
M.Gillon$^{6,4}$,
\newauthor 
L.Hebb$^{7}$, 
M.Lendl$^{4}$,
N.Parley$^{1}$,
F.Pepe$^{4}$, 
D.Pollacco$^{8}$, 
D.Segransan$^{4}$, 
E.Simpson$^{8}$, 
\newauthor
R.A.Street$^{3}$ 
and
S.Udry$^{4}$
\\
$^{1}$SUPA, School of Physics and Astronomy, University of St. Andrews, North Haugh, St Andrews, KY16 9SS. \\
$^{2}$Astrophysics Group, Keele University, Staffordshire, ST5 5BG, UK. \\
$^{3}$Las Cumbres Observatory, 6740 Cortona Drive Suite 102, Goleta, CA 93117, USA. \\
$^{4}$Observatoire astronomique de l`Universit\'en de Gen\'eve, 51 Chemin des Maillettes, 1290 Sauverny, Switzerland. \\
$^{5}$Department of Physics and Astronomy, University of Leicester, Leicester, LE1 7RH, UK. \\
$^{6}$Institut d`Astrophysique et de G\'eophysique, Universit\'e de Li\'ege, All\'ee de 6 Ao\^ut, 17, Bat B5C, Li\'ege 1, Belgium. \\
$^{7}$Vanderbilt University, Department of Physics and Astronomy, Nashville, TN 37235. \\
$^{8}$Astrophysics Research Centre, School of Mathematics \& Physics, Queen`s University, University Road, Belfast, BT7 1NN, UK. \\
}



\date{Received / Accepted}

\pagerange{\pageref{firstpage}--\pageref{lastpage}} \pubyear{2010}

\maketitle

\label{firstpage}

\begin{abstract}
We report the detection of a 0.6~M$_J$ extrasolar planet by WASP-South, WASP-25b, transiting its solar-type host star every 3.76~days. A simultaneous analysis of the WASP, FTS and Euler photometry and CORALIE spectroscopy yields a planet of R$_p$~=~1.22~$R_J$ and M$_p$~=~0.58~$M_J$ around a slightly metal-poor solar-type host star, $[$Fe/H$]$~=~-0.05$\pm$0.10, of R$_{\ast}$~=~0.92~$R_{\odot}$ and M$_{\ast}$~=~1.00~$M_{\odot}$. WASP-25b is found to have a density of $\rho_p~=~0.32~\rho_J$, a low value for a sub-Jupiter mass planet. We investigate the relationship of planetary radius to planetary equilibrium temperature and host star metallicity for transiting exoplanets with a similar mass to WASP-25b, finding that these two parameters explain the radii of most low-mass planets well. 
\end{abstract}



\section{Introduction}

To date, over 440 exoplanets have been discovered, including more than 70 detected by the transit method\footnotemark \footnotetext[1]{www.exoplanet.eu}. The transit method together with follow-up radial velocity observations allow measurement of both the mass and radius of the planet, leading to a value for the planet`s bulk density \citep{char00}. The atmospheric composition of transiting exoplanets can also be investigated through high-precision photometric and spectroscopic measurements, see e.g. \citet{char02}.

A wide range of transiting exoplanets radii has been found and there has been much investigation into the factors that may influence a planet`s radius. For example, \citet{guillot06} propose a negative relationship between the metallicity of a host star and the radius of an orbiting planet, caused by an increase in the amount of heavy elements in the planet, leading to a more massive core and hence smaller radius for a given mass. Alternatively, \citet{burrows07} consider that increasing the metallicity may increase the opacity of an exoplanet`s atmosphere, retarding cooling and leading to a larger radius for a given mass. Another influence on a planet`s radius may be the equilibrium temperature of the planet \citep{guillot02}, determined by the stellar irradiation and the planet`s distance from its host star. Tidal heating due to the circularisation of the orbits of close-in exoplanets may also play a role in inflating the planetary radius \citep{bodenheimer03,jackson08}. One motivation of the SuperWASP project is to detect enough transiting exoplanets, with a wide range of orbital and compositional parameters, to allow analyses that may distinguish between such differing models. 


In this paper, we report the discovery of a 0.6~$M_J$ planet orbiting a solar-mass star, WASP-25 (=TYC6706-861-1, =1SWASP J130126.36-273120.0), in the southern hemisphere. Analysis of photometric and spectroscopic data reveals WASP-25b to be another low-density planet, comparable to HD~209458b \citep{char00}. We also analyse the dependence of the radii of low-mass planets on host star metallicity and planetary equilibrium temperature, including WASP-25b and 18 other transiting planets, finding a relationship using Singular Value Decomposition analysis that gives an excellent agreement between observed and calibrated radii.

In Section \ref{obs} we describe the photometric and spectrscopic observations and data reduction procedures. In Section \ref{params} we present the stellar and planetary parameters extracted from these data. Finally, in Section \ref{disc} we compare WASP-25b with the ensemble of known planets of similar mass, and examine the relationship between stellar metallicity, irradiating flux (via the planet's equilibrium temperature) and planet radius. 

\section{Observations}
\label{obs}

\subsection{Photometric Obervations}

The WASP-South observatory is located at SAAO in South Africa, and consists of eight 11cm telescopes of $7.8^{\circ} \times 7.8^{\circ}$ field of view each, on a single fork mount. The cameras scan repeatedly through eight to ten sets of fields, taking 30 second exposures. See \citet{pollacco06} for more details on the WASP project and the data reduction procedure, and \citet{cameron07} and \citet{pollacco08} for an explanation of the candidate selection process.

\begin{figure}
\includegraphics[width=80mm]{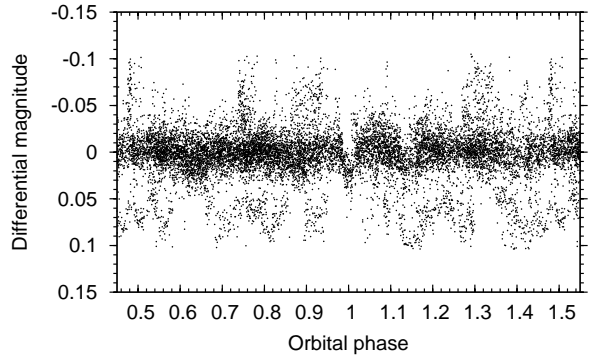}
\caption{WASP discovery lightcurve folded on the orbital period of $P$~=~3.765 d.
Points with error above $3\times median$  were clipped, where $median$~=~0.012 mag}
\label{fig:waspphot}
\end{figure}

WASP-25 was observed by WASP-South in 2006, 2007 and 2008, producing a total of 14,186 photometric datapoints. The 2007 dataset showed the transit event most clearly, detected at a period of 3.76 days. Figure \ref{fig:waspphot} shows the WASP-South discovery lightcurve, using data from all seasons.

Further photometric observations were subsequently obtained on 3 April 2010 using the LCOGT 2m Faulkes Telescope South (FTS) at Siding Spring, Australia. Observations were obtained using the fs01 Spectral camera containing a $4096 \times 4096$ pixel Fairchild CCD which was binned $2\times 2$ giving 0.303 arcsec/pixel and a field of view of $\sim10\arcmin\times10\arcmin$. 285 datapoints were obtained through a Pan-STARRS z filter, capturing an entire transit.

The data were pre-processed through the WASP Pipeline \citep{pollacco06} to perform masterbias/flat creation, debiassing and flatfield correction in the standard manner. Object detection and aperture photometry were performed using the DAOPHOT \citep{stetson87} package within the IRAF environment\footnote{IRAF is distributed by the National Optical Astronomy Observatory, which is operated by the Association of Universities for Research in Astronomy (AURA) under cooperative agreement with the National Science Foundation.} with an aperture size of 18 binned pixels. Differential magnitudes of WASP-25 were formed by a weighted combination of the flux relative to 26 comparison stars within the field of view

\begin{figure}
\includegraphics[angle=0,width=80mm]{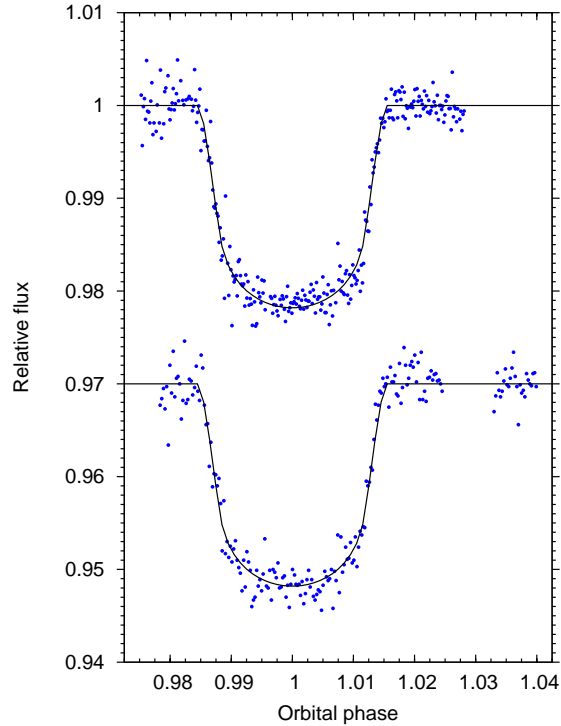}
\caption{Top lightcurve: FTS follow-up photometry of WASP-25 on 3 April 2010. The central transit time is at HJD = 5290.05617. Bottom lightcurve: Euler photometry on 10 April 2010 with central transit time at HJD = 5297.58552.}
\label{fig:ftsphot}
\end{figure}

A second full transit was observed a week later on 10 April 2010 in the R-band with EulerCam on the 1.2m telescope at La Silla, Chile, comprising 204 datapoints. The resulting lightcurves are shown in Figure \ref{fig:ftsphot}.

 
\subsection{Spectroscopic Observations}

\begin{figure}
\includegraphics[width=80mm]{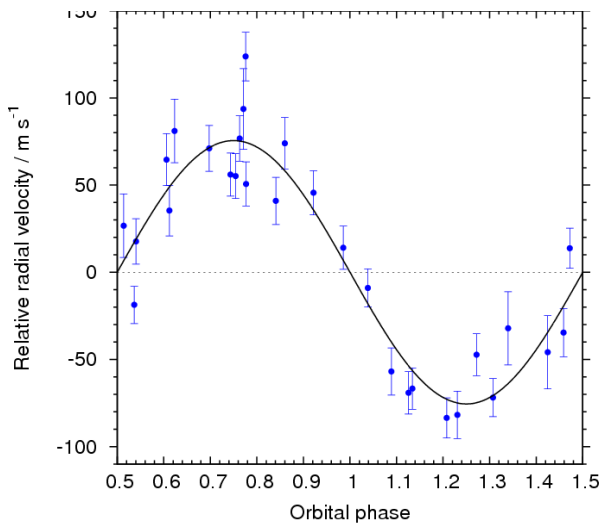}
\caption{Top plot: Radial velocity measurements. The solid line is the best-fitting MCMC solution. The centre-of-mass velocity, $\gamma$ = -2.6323 km s$^{-1}$, was subtracted.}
\label{fig:rv}
\end{figure}

\begin{figure}
\includegraphics[angle=90,width=80mm]{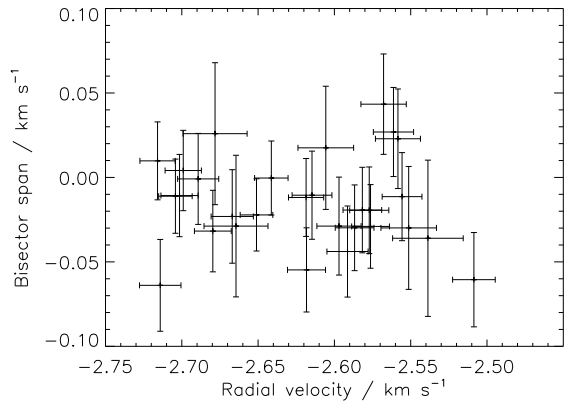}
\caption{Bisector spans versus radial velocity, where bisector uncertainties are taken to be equal to twice the radial velocity uncertainties.}
\label{fig:bis}
\end{figure}

WASP-25, a V$_{mag}$ = 11.9 star, was observed 29 times with the CORALIE spectrograph on the 1.2m Euler telescope, between 29 December 2008 and 28 June 2009. The spectra were processed using the standard data reduction pipeline for CORALIE \citep{baranne96,mayor09}, plus a correction for the blaze function \citep{triaud10}. These data are given in Table \ref{tab:coralie} and shown phase-folded in Figure \ref{fig:rv}: the low-amplitude radial velocity variations and the lack of correlation between the bisector spans and radial velocity \citep{queloz01}, shown in Figure \ref{fig:bis}, are consistent with a planet-mass object orbiting the host star. 

\begin{table}
\caption{CORALIE radial velocity measurements of WASP-25}
\label{tab:coralie}
\begin{tabular*}{0.5\textwidth}{@{\extracolsep{\fill}}cccc}
\hline
BJD--2\,400\,000 & RV & $\sigma$$_{\rm RV}$ & BS\\
 & (km s$^{-1}$) & (km s$^{-1}$) & (km s$^{-1}$)\\
\hline
54829.8227 & -2.577 & 0.013 & -0.019\\
54896.7698 & -2.651 & 0.011 & -0.022\\
54940.7092 & -2.716 & 0.012 & ~0.010\\
54941.7043 & -2.619 & 0.012 & -0.012\\
54942.7257 & -2.576 & 0.012 & -0.029\\
54943.6374 & -2.618 & 0.012 & -0.055\\
54944.7155 & -2.680 & 0.012 & -0.032\\
54945.7265 & -2.615 & 0.013 & -0.011\\
54946.6166 & -2.582 & 0.013 & -0.019\\
54947.6016 & -2.641 & 0.011 & -0.000\\
54947.7912 & -2.689 & 0.013 & ~0.001\\
54948.6130 & -2.704 & 0.011 & -0.011\\
54949.8031 & -2.551 & 0.018 & -0.030\\
54950.6221 & -2.591 & 0.013 & -0.044\\
54951.6953 & -2.701 & 0.012 & -0.011\\
54971.6453 & -2.678 & 0.021 & ~0.026\\
54972.6724 & -2.561 & 0.013 & ~0.027\\
54973.5157 & -2.587 & 0.013 & -0.030\\
54974.6787 & -2.714 & 0.014 & -0.064\\
54975.5379 & -2.667 & 0.014 & -0.023\\
54976.6837 & -2.556 & 0.013 & -0.011\\
54982.6194 & -2.664 & 0.021 & -0.029\\
54983.6213 & -2.568 & 0.015 & ~0.043\\
54983.6446 & -2.597 & 0.015 & -0.029\\
54984.5785 & -2.558 & 0.015 & ~0.023\\
54985.6100 & -2.699 & 0.012 & ~0.004\\
54995.5555 & -2.509 & 0.014 & -0.061\\
55009.6287 & -2.606 & 0.018 & ~0.018\\
55010.5967 & -2.539 & 0.023 & -0.036\\
\hline
\end{tabular*}
\end{table}


\section{System Parameters}
\label{params}

\subsection{Stellar Parameters}

The CORALIE spectra were placed on a common wavelength scale and co-added to produce a higher signal-to-noise spectrum, allowing an analysis of the host star and thus measurement of the stellar temperature, gravity, metallicity, $v$ sin $i$ and elemental abundances, given in Table \ref{tab:specanalysis}, where $\eta$ is microturbulence. The spectral data were analysed using the UCLSYN spectral synthesis package \citep{smalley01} and ATLAS9 models without convective overshooting \citep{castelli97}. The effective temperature and log~g were determined using the H$\alpha$ and H$\beta$ lines, and the Na ID and Mg Ib lines respectively, while the Ca H and K lines provided a check on those values. Further details of the spectral analysis are given in \citet{west09}. The analysis yielded T$_{\mbox{eff}} = 5750\pm100$K and $[$Fe/H$]$ = -0.05$\pm$0.10.

\begin{table}[!h]
\begin{center}
\caption{Stellar Parameters of WASP-25}
\label{tab:specanalysis}
\begin{tabular}{cc}
\hline
Parameter & Value \\
\hline
T$_{\mbox{eff}}$ & 5750 $\pm$ 100 K \\
log $g$ & 4.5 $\pm$ 0.15 \\
$\eta$ & 1.1 $\pm$ 0.1 km s$^{-1}$ \\
$v$ sin $i$ & 3.0 $\pm$ 1.0 km s$^{-1}$ \\
Spectral Type & G4 \\
 & \\
$[$Fe/H$]$ & -0.05 $\pm$ 0.10 \\
$[$Si/H$]$ & 0.00 $\pm$ 0.06 \\
$[$Ca/H$]$ & 0.08 $\pm$ 0.14 \\
$[$Ti/H$]$ & 0.04 $\pm$ 0.07 \\
$[$Ni/H$]$ & -0.08 $\pm$ 0.10 \\
log[Li/H] & 1.63 $\pm$ 0.09 \\
 & \\
V (mag) & 11.88 \\
\hline
\multicolumn{2}{c} 1SWASP J130126.36-273120.0 \\
\hline
\end{tabular}
\end{center}
\end{table}

\subsection{System Parameters}

The WASP-South, FTS and Euler photometry were simultaneously analysed with the CORALIE radial velocity data in a Markov-Chain Monte-Carlo (MCMC) analysis. This analysis is described in \citet{cameron07}, but is here modified as described in \citet{enoch10} to determine stellar mass using a calibration on T$_{\mbox{eff}}$, log $\rho$ and [Fe/H] (similar to the T$_{\mbox{eff}}$, log $g$ and [Fe/H] calibration described in \citet{torres09}). The temperature and metallicity were obtained through spectral analysis (given in Table \ref{tab:specanalysis}), and the density is determined directly from the photometry. Mass and radius values obtained for other WASP host stars via this method agree closely with values obtained through isochrone analysis. 

An initial analysis was performed allowing the eccentricity value to float, producing a value of $e = 0.123^{+0.041}_{-0.046}$. This eccentricity value is consistent with 0 at the $3\sigma$ level, and was suspected to not be significant, as can occur from spurious asymmetries in quadrature fits due to noise \citep{laughlin05}, so a second analysis was performed with eccentricity fixed to 0. We performed an F-test on the radial velocity residuals from the circular and floating eccentricity fits, resulting in a value of 0.897 which shows that the eccentric fit is not significant. 

The resulting circular model best-fit parameters for the star-planet system are listed in Table \ref{tab:results}, using the best-fit parameters from the $e=0$ fit and the uncertainties from the fit allowing eccentricity to float, to sufficiently account for uncertainty due to unknown eccentricity. The results show WASP-25 to be a solar analogue of one solar mass and $0.92~\pm~0.04$ solar radius, and WASP-25b to be a bloated hot Jupiter of $0.58~\pm~0.04~M_J$ and $1.22~\pm~0.06~R_J$, giving a planet density of $\rho = 0.32^{+0.04}_{-0.03}~\rho_J$. 

\begin{figure}
\includegraphics[angle=270,width=80mm]{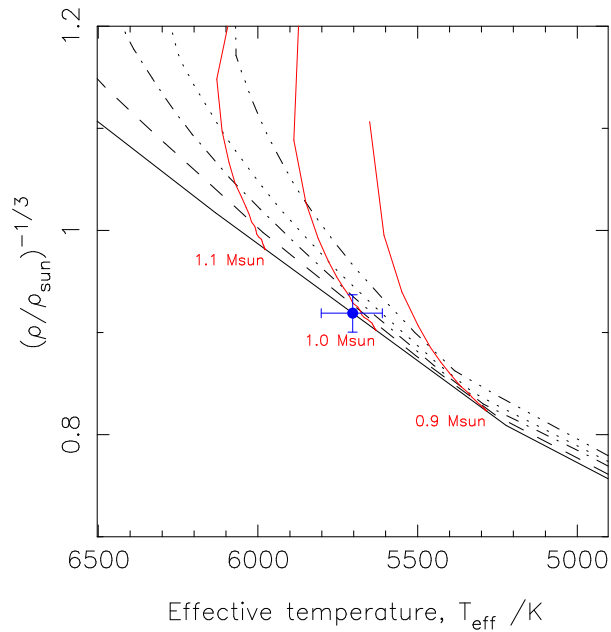}
\caption{Isochrone tracks from \citet{girardi00} for WASP-25 using the best-fit metallicity of -0.07 and stellar density $1.29\rho_{\odot}$. Isochrones are:  0.01995 (solid),  0.7943 (dashed),  1.99 (dot-dash),  3.16 (dotted),  5.01 (triple dot-dash) Gyr. }
\label{fig:iso}
\end{figure}

The stellar density of $1.29\pm-0.11~\rho_{\odot}$ obtained from the MCMC analysis was used along with the determined stellar temperature and metallicity values in an interpolation of the \citet{girardi00} stellar evolution tracks, see Figure \ref{fig:iso}. Using the best-fit metallicity of $-0.07$ indicates that WASP-25 has a mass of $1.02^{+0.03}_{-0.07}M_{\odot}$, agreeing well with the calibrated MCMC result of $1.00\pm0.03M_{\odot}$, and an age of $0.02^{+3.96}_{-0.01}$~Gyr.


\begin{table*}
\begin{center}
\setlength{\extrarowheight}{5pt}
\caption{System Parameters of WASP-25}
\label{tab:results}
\begin{tabular}{lcr}
\hline
Parameter & Symbol & Value  \\
\hline
Period (days) & $P$ & $3.764825~\pm~0.000005$ \\
Transit Epoch (HJD) & $T_0$ & $5274.99649~\pm~0.00017$ \\
Transit duration (days) & $D$ & $0.116~\pm~0.001$ \\
Planet/Star area ratio & $R_p^2/R_{\ast}^2$ & $0.0187~\pm~0.0002$ \\
Impact Parameter & $b$ & $0.38^{+0.06}_{-0.07}$ \\
 & \\
Stellar Reflex Velocity (ms$^{-1}$) & $K_1$ & $75.5~\pm~5.3$ \\
Centre-of-mass Velocity (ms$^{-1}$) & $\gamma$ & $-2632.3~\pm~0.6$ \\
Orbital separation (AU) & $a$ & $0.0473~\pm~0.0004$ \\
Orbital inclination (deg) & $i$ & $88.0~\pm~0.5$ \\
Orbital eccentricity & $e$ & 0 (adopted) \\
 & \\
Stellar mass (M$_{\odot}$) & $M_{\ast}$ & $1.00~\pm~0.03$ \\
Stellar radius (R$_{\odot}$) & $R_{\ast}$ & $0.92~\pm~0.04$ \\
Stellar surface gravity (log $g_{\odot}$) & log $g_{\ast}$ & $4.51~\pm~0.03$ \\
Stellar density ($\rho_{\odot}$) & $\rho_{\ast}$ & $1.29~\pm~0.10$ \\
Stellar metallicity & $[$Fe/H$]$ & $-0.07~\pm~0.10$ \\
Stellar effective temperature & T$_{eff}$ & $5703~\pm~100$ \\
 & \\
Planet mass (M$_J$) & $M_p$ & $0.58~\pm~0.04$ \\
Planet radius (R$_J$) & $R_p$ & $1.22^{+0.06}_{-0.05}$ \\
Planet surface gravity (log $g_J$) & log $g_p$ & $2.95~\pm~0.04$ \\
Planet density ($\rho_J$) & $\rho_p$ & $0.317^{+0.036}_{-0.031}$ \\
Planet temperature (A=0, F=1) (K) & T$_{\mbox{eq}}$ & $1212~\pm~35$ \\
\hline
\end{tabular}
\end{center}
\end{table*}




\section{Discussion}
\label{disc}

A density of $0.32^{+0.04}_{-0.03}~\rho_J$ places WASP-25b amongst the bloated hot Jupiters, with over 80\% of known transiting exoplanets being more dense\footnotemark \footnotetext[3]{www.exoplanet.eu}. \citet{guillot06} proposed a correlation between the metallicity of a host star and the amount of heavy elements present in the planet. A larger amount of heavy elements is likely to produce a more massive rocky core and hence would lead to larger planetary radii for lower metallicity stars, for a given planetary mass \citep{fressin07}. Alternatively, the heavy elements could increase the opacity of the planetary interior, potentially increasing the radius \citep{burrows07}. 

WASP-25b has a mass close to the value above which the gravitational, potential energy begins to exceed the electrical binding energy of electrons in atoms as the main force opposing electron degeneracy pressure in the deep planetary interior \citep{lynden01}. It is in this lower mass range that we see the greatest range of radii at a given mass. This suggests that environmental and compositional differences between individual planets have the greatest influence on their radii in this mass range. 

\begin{figure}
\includegraphics[angle=90,width=80mm]{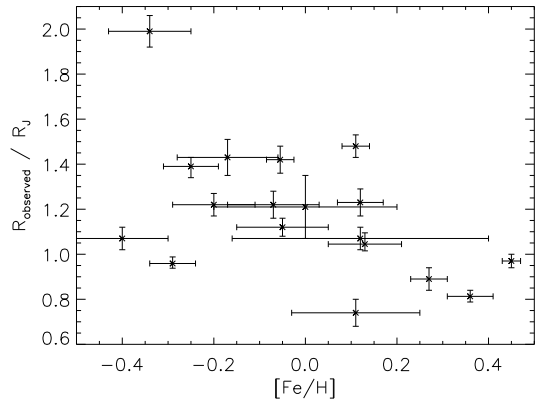}
\caption{Radius versus metallicity of 18 planets of 0.1-0.6~$M_J$.}
\label{fig:radmet}
\end{figure}

To investigate this relationship, we calculated the correlation between radius and metallicity for all known transiting exoplanets for which metallicity data was available (67 planets), finding only a very weak relationship with a correlation coefficient of -0.17. However, splitting the planets into low-mass ($0.1-0.6~M_J$) and higher-mass ($>0.6~M_J$) sets resulted in correlation coefficients of -0.53 (with a probability of no correlation of 0.022) and 0.07 respectively, indicating that metallicity plays a role in determining the radii of low-mass planets but not those of higher mass. We plot the radii against host star metallicity of low-mass planets in Figure~\ref{fig:radmet}. We concentrate on the set of 18 low-mass planets, of which 9 are WASP planets, including WASP-25b, see Table \ref{tablow}), in further investigating planetary radii.

\begin{figure}
\includegraphics[angle=90,width=80mm]{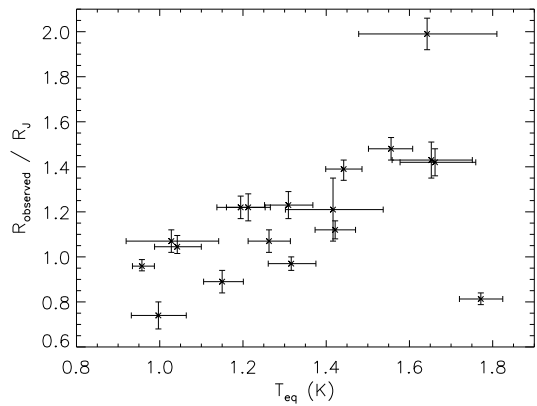}
\caption{Radius versus equilibrium temperature of 18 planets of 0.1-0.6~$M_J$.}
\label{fig:radteq}
\end{figure}

A major influence on the radius of an exoplanet is likely to be the planet's equilibrium temperature \citep{fressin07}, T$_{\mbox{eq}}$, defined as 
\begin{equation}
\mbox{T}_{\mbox{eq}} = \mbox{T}_{\ast,\mbox{eff}} \left( \frac{1-A}{F} \right) \sqrt{\frac{ R_{\ast}}{2 a}}
\end{equation}
\noindent where T$_{\ast,\mbox{eff}}$ is the host star's effective temperature, $R_{\ast}$ is the stellar radius, $a$ is the semi-major axis, $A$ is the planetary Bond albedo and $F$ is the fraction of the planet re-radiating flux. Jupiter has a Bond albedo of 0.28 \citep{taylor65} but albedo values have not so far been determined for the majority of exoplanets, and \citet{rowe08} found a very low albedo for HD~209458b, at just $0.04\pm0.05$. We set $A=0$ ($F=1$) here to calculate the equilibrium temperatures. Not knowing the true albedo values for all the planets studied here produces an uncertainty in the values calculated for equilibrium temperature of $(1-A)^{1/4}$. We find the correlation coefficient between the (approximate) equilibrium temperature and radius of all 18 planets to be 0.55, a moderate relationship, shown in Figure \ref{fig:radteq}. The correlation of radii and equilibrium temperature is stronger than this for the majority of planets - removing the two outliers, WASP-17b and HD~149026b, from the correlation produces a coefficient of 0.82 for the remaining 16 planets, a strong relationship. Accurate albedo values, and hence equilibrium temperatures, may produce an even stronger correlation than found here.


\begin{figure}
\includegraphics[angle=90,width=80mm]{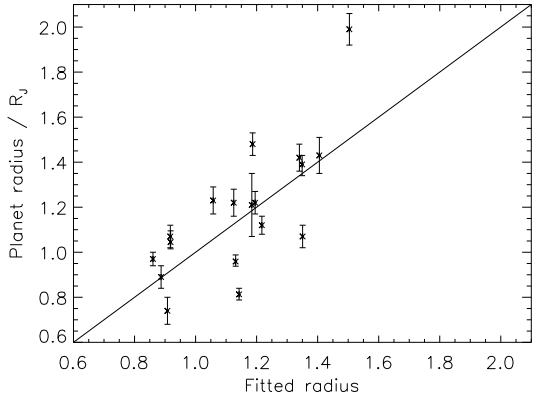}
\caption{Results of SVD fit on the radii of 18 planets of 0.1-0.6~$M_J$.}
\label{fig:svdrad}
\end{figure}

\begin{figure}
\includegraphics[angle=90,width=80mm]{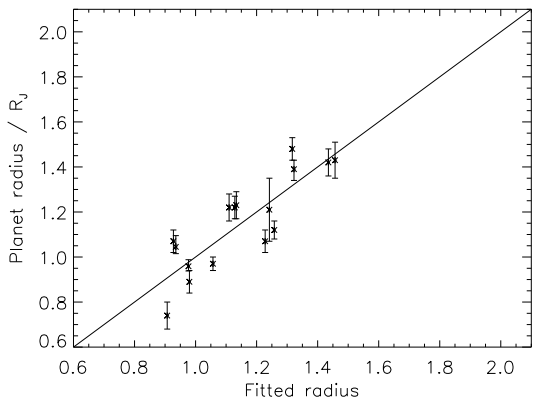}
\caption{Results of SVD fit on the radii of planets of 0.1-0.6~$M_J$, excluding WASP-17b and HD~149026b.}
\label{fig:svdnoout}
\end{figure}

Since both the host star metallicity and the planet`s equilibrium temperature seem to explain the radii of low-mass planets in part, we performed a weighted Singular Value Decomposition (SVD) fit to fully quantify the relationship. This resulted in the equation
\begin{equation}
R_p = 0.48 + 0.50 (\mbox{T}_{\mbox{eq}}/1000K) - 0.61 [\mbox{Fe/H}]
\end{equation}
\noindent producing the fit to the planetary radii shown in Figure \ref{fig:svdrad}. Removing the two outliers mentioned above, WASP-17b and HD~149026b, leads to a much better fit, shown in Figure \ref{fig:svdnoout}, implying that additional parameters particularly affect these two planets. The SVD fit for the 16 planets is 
\begin{equation}
R_p = 0.20 + 0.73 (\mbox{T}_{\mbox{eq}}/1000K) - 0.25 [\mbox{Fe/H}]
\end{equation}

The negative slope of the metallicity correlation and of the SVD coefficient implies that the effect of a massive core \citep{guillot06} outweighs the effect of increased opacity in the interior of the planet, i.e. increased metallicity leads to a smaller planetary radius.




\begin{table*}[!h]
\begin{center}
\setlength{\extrarowheight}{5pt}
\caption{Details of the 18 low-mass planets.}
\label{tablow}
\begin{tabular}{lrrrrrrrl}
\hline
Planet & $M_p/M_J$ & $R_p/R_J$ & $a/AU$ & T$_{\ast,\mbox{eff}}$(K) & $[$Fe/H$]$ & $R_{\ast}/R_{\odot}$ & T$_{\mbox{eq}}$(K) & Reference\\
\hline
HAT-12b & 0.21 & $0.96^{+0.03}_{-0.02}$ & $0.0384\pm0.0003$ & $4650\pm60$ & $-0.20\pm0.05$ & $0.70^{+0.02}_{-0.01}$ & $957^{+30}_{-23}$ & \citet{hartman09} \\
WASP-29b & 0.25 & $0.74\pm0.06$ & $0.0456\pm0.0006$ & $4800\pm150$ & $0.11\pm0.14$ & $0.85\pm0.05$ & $997^{+67}_{-65}$ & \citet{hellier10} \\
WASP-21b & 0.30 & $1.07\pm0.05$ & $0.0520\pm0.0004$ & $5800\pm100$ & $-0.4\pm0.1$ & $1.01^{+0.02}_{-0.03}$ & $1263^{+51}_{50}$ & \citet{bouchy10} \\
HD149026b & 0.37 & $0.81\pm0.03$ & $0.0431^{+0.0007}_{-0.0006}$ & $6147\pm50$ & $0.36\pm0.05$ & $1.54^{+0.05}_{-0.04}$ & $1772^{+53}_{-52}$  & \citet{sato05},\citet{carter09} \\
Kepler-7b & 0.43 & $1.48^{+0.05}_{-0.05}$ & $0.0622^{+0.0011}_{-0.0008}$ & $5933\pm44$ & $0.11\pm0.03$ & $1.84^{+0.07}_{-0.07}$ & $1556^{+52}_{-54}$  &  \citet{latham10} \\
WASP-11b & 0.46 & $1.045^{+0.05}_{-0.03}$ & $0.043^{+0.002}_{-0.002}$ & $4980\pm60$ & $0.13\pm0.08$ & $0.81^{+0.03}_{0.03}$ & $1042^{+58}_{-55}$ & \citet{west09b} \\
WASP-13b & 0.46 & $1.21^{+0.14}_{-0.14}$ & $0.0527^{+0.0017}_{-0.0019}$ & $5826\pm100$ & $0.0\pm0.2$ & $1.34^{+0.13}_{-0.13}$ & $1416^{+121}_{-114}$ & \citet{skillen09} \\
CoRoT-5b & 0.47 & $1.39^{+0.04}_{-0.05}$ & $0.0495^{+0.0003}_{-0.0003}$ & $6100\pm65$ & $-0.25\pm0.06$ & $1.19^{+0.04}_{-0.04}$ & $1442^{+44}_{-43}$ & \citet{rauer09} \\
WASP-17b & 0.49 & $1.99\pm0.07$ & $0.0518^{+0.0017}_{-0.0018}$ & $6600\pm100$ & $-0.34\pm0.09$ & $1.38^{+0.19}_{-0.19}$ & $1645^{+169}_{-166}$ & \citet{anderson09} \\
WASP-6b & 0.50 & $1.22^{+0.05}_{-0.05}$ & $0.0421^{+0.0008}_{-0.0013}$ & $5450\pm100$ & $-0.20\pm0.09$ & $0.87^{+0.03}_{-0.04}$ & $1195^{+59}_{-57}$ & \citet{gillon09} \\
HAT-1b & 0.52 & $1.23^{+0.06}_{-0.06}$ & $0.0553^{+0.0014}_{-0.0014}$ & $6047\pm56$ & $0.12\pm0.05$ & $1.12^{+0.05}_{-0.05}$ & $1309^{+59}_{-57}$ & \citet{bakos07} \\
OGLE-111b & 0.53 & $1.07^{+0.05}_{-0.05}$ & $0.047^{+0.001}_{-0.001}$ & $5070\pm400$ & $0.12\pm0.28$ & $0.83^{+0.03}_{-0.03}$ & $1028^{+114}_{-109}$ & \citet{pont04} \\
WASP-15b & 0.54 & $1.43^{+0.08}_{-0.08}$ & $0.0499^{+0.0018}_{-0.0018}$ & $6300\pm100$ & $-0.17\pm0.11$ & $1.48^{+0.07}_{-0.07}$ & $1653^{+99}_{-94}$ & \citet{west09} \\
WASP-22b & 0.56 & $1.12^{+0.04}_{-0.04}$ & $0.0468^{+0.0004}_{-0.0004}$ & $6000\pm100$ & $-0.05\pm0.1$ & $1.13^{+0.03}_{-0.03}$ & $1421^{+49}_{-48}$ & \citet{maxted10} \\
XO-2b & 0.57 & $0.97^{+0.03}_{-0.03}$ & $0.037^{+0.002}_{-0.002}$ & $5340\pm32$ & $0.45\pm0.02$ & $0.96^{+0.02}_{-0.02}$ & $1315^{+59}_{-55}$ & \citet{mccull06} \\
WASP-25b & 0.58 & $1.26^{+0.06}_{-0.06}$ & $0.0474^{+0.0004}_{-0.0004}$ & $5750\pm100$ & $-0.05\pm0.02$ & $0.95^{+0.04}_{-0.04}$ & $1241^{+55}_{-55}$ & - \\
HAT-3b & 0.60 & $0.89^{+0.05}_{-0.05}$ & $0.0389^{+0.0007}_{-0.0007}$ & $5185\pm46$ & $0.27\pm0.04$ & $0.82^{+0.04}_{-0.04}$ & $1150^{+51}_{-45}$ & \citet{torres07} \\
Kepler-8b & 0.60 & $1.42^{+0.06}_{-0.06}$ & $0.0483^{+0.0006}_{-0.0012}$ & $6213\pm150$ & $-0.055\pm0.03$ & $1.49^{+0.06}_{-0.06}$ & $1662^{+98}_{-84}$ & \citet{jenkins10} \\
\hline
\end{tabular}
\end{center}
\end{table*}



\section{Conclusions}

We have reported the detection of a 0.58~$M_J$ planet, WASP-25b, transiting a slightly metal-poor solar-mass star in the southern hemisphere with an orbital period of 3.76~days. WASP-25b has a low density, $0.29~\rho_J$, and we investigate its bloated radius, $R_p~=~1.26~R_J$. We find that the radii of most transiting exoplanets of a similar mass to WASP-25b can be explained well by a calibration to the host star metallicity and planetary equilibrium temperature.

\section*{Acknowledgements}
\footnotesize

WASP-South is hosted by the South African Astronomical Observatory and we are grateful for their ongoing support and assistance. Funding the WASP comes from consortium universities and from the UK`s Science and Technology Facilities Council.

\normalsize


\bibliographystyle{mn2e}
\bibliography{planet.bib}

\label{lastpage}

\end{document}